\documentclass[a4paper,12pt]{article}
\usepackage{amsmath}
\usepackage{amssymb}

 \newread\testifexists
\def\GetIfExists #1 {\immediate\openin\testifexists=#1
    \ifeof\testifexists\immediate\closein\testifexists\else
    \immediate\closein\testifexists\input #1\fi}

\immediate\openin\testifexists=e
    \ifeof\testifexists\immediate\closein\testifexists\else
    \immediate\closein\testifexists\input e\fipsf

\begin{document}

\title{ $B^0  \to \phi\phi$ Decay
in Perturbative QCD Approach}
\author{Cai-Dian L\"u$^{a,b}$, \,  Yue-long  Shen$^b$ ,\,   Jin Zhu$^b$\\
{\it \small $a$ CCAST (World Laboratory), P.O. Box 8730,
   Beijing 100080, China}\\
{\it \small $b$  Institute of High Energy Physics, CAS, P.O.Box
918(4) } {\it \small Beijing 100049, China\footnote{Mailing
address.}}}

\maketitle

\begin{abstract}
The rare decay $B^0  \to \phi\phi$ can occur only via penguin
annihilation
 topology in the standard model. We calculate this channel in the
 perturbative QCD approach. The predicted
 branching ratio is very small at ($10^{-8}$). We also  give the polarization
 fractions, which shows that the transverse polarization
 contribution is comparable to the longitudinal
 one, due to a big transverse contribution from factorizable diagrams.
 The small branching ratio in SM, makes it sensitive
 to any new physics contributions.
\end{abstract}

\section{Introduction}
\hspace*{\parindent} The study of B meson decays has offered a
good place to test the standard model (SM) and to give some
important constraints on the SM parameters. Recently, more
attentions have been paid to the $B\rightarrow VV$ decay modes.
The transverse polarization of the vector meson can contribute to
the decay width, and the fraction of each kind of polarization has
been or will be measured. In some penguin dominated decay modes,
such as $B\rightarrow \phi K^\ast$ \cite{1}, the experimental
results for polarization differ from most theoretical predictions
\cite{2}, which has been considered as a puzzle and lots of
discussions have been given \cite{5,7}. So the polarization
problem in the $B\rightarrow VV$ decay modes brings a new
challenge to the standard model, maybe it is a signal of new
physics \cite{7,8}.

  In this work we will calculate the branching ratio and the
polarization fractions of the charmless decay channel $B^0
\rightarrow \phi \phi$ with perturbative QCD  approach (PQCD)
\cite{9,10}. In this channel, the initial $\bar{b}$ quark and the
light valence $d$ quark in the $B$ meson don't appear in the final
states, so it must be an annihilation topology in Feynman
diagrams. Annihilation diagrams can't be calculated in
factorization approach \cite{12,13} or in QCD improved
factorization approach \cite{14} for its endpoint singularity, but
in PQCD approach this singularity can be regulated by Sudakov form
factor and threshold resummation, so the PQCD calculations can
give converging results and have prediction power. In this
channel, since no tree level operators can contribute, the
dominant contribution comes from  penguin operators. The
annihilation topology is usually suppressed relative to the
emission topology which can appear in other modes, so this channel
is a rare decay mode, and hasn't been measured in the $B$
factories.

In the next section   we   give our theoretical formulae based on
the PQCD framework. Then we show the numerical results and a brief
conclusion in the third section.

\section{Perturbative calculation}

\hspace*{\parindent} For simplicity, we work in the B meson rest
frame, and   adopt the light-cone coordinate system. Then the
four-momentum of the B meson and the two $\phi$ mesons in the
final state can be written as:
\begin{eqnarray}
\nonumber &&P_1=\frac{M_B}{\sqrt{2}}(1,1,{\bf0_T}),\\
\nonumber &&P_2=\frac{M_B}{\sqrt{2}}(1-r,r,{\bf 0_T}),\\
&&P_3=\frac{M_B}{\sqrt{2}}(r,1-r,{\bf0_T}),
\end{eqnarray}
in which $r$ is defined by
$r=\frac{1}{2}(1-\sqrt{1-4M_{\phi}^2/M_{B}^2})\simeq
M_{\phi}^2/M_{B}^2\ll 1$. To extract the helicity amplitudes, we
should parameterize the polarization vectors. The longitudinal
polarization vector must satisfy the orthogonality and
normalization: ${\epsilon_{2L} \cdot P_2}=0,\,\,{ \epsilon_{3L}
\cdot P_3 }=0 $, and ${ \epsilon_{2L}}^2={ \epsilon_{3L}}^2=-1$.
Then we can give the manifest form as follows:
\begin{eqnarray}
\nonumber{ \epsilon_{2L}}=\frac{1}{\sqrt{2r}}(1-r,-r,{\bf0_T}),\\
{ \epsilon_{3L}}=\frac{1}{\sqrt{2r}}(-r,1-r,{\bf0_T}).
\end{eqnarray}
As to the transverse polarization vectors, we can choose the
simple form:
\begin{eqnarray}
\nonumber{ \epsilon_{2T}}=\frac{1}{\sqrt{2}}(0,0,{\bf1_T}),\\
{ \epsilon_{3T}}=\frac{1}{\sqrt{2}}(0,0,{\bf1_T}).
\end{eqnarray}

 Only penguin operators can contribute to this decay channel, so
the relevant effective weak Hamiltonian can be written as
\cite{15}:
\begin{eqnarray}
{\cal
H}_{eff}=\frac{G_F}{\sqrt{2}}V_{tb}V_{td}^*C_i(\mu)O_i(\mu),\,\,\,\,i=3-10,
\end{eqnarray}
where $C_i$ are QCD corrected Wilson coefficients, and $O_i$ are
the usual penguin operators with the form
\begin{equation}
\begin{array}{ll}
O_3=(\bar{s}_ib_i)_{V-A}\sum\limits_q(\bar{q}_jq_j)_{V-A},&
O_4=(\bar{s}_ib_j)_{V-A}\sum\limits_q(\bar{q}_jq_i)_{V-A} \\
O_5=(\bar{s}_ib_i)_{V-A}\sum\limits_q(\bar{q}_jq_j)_{V+A},&
 O_6=(\bar{s}_ib_j)_{V-A}\sum\limits_q(\bar{q}_jq_i)_{V-A},\\
 O_7=\frac{3}{2}(\bar{s}_ib_i)_{V-A}\sum\limits_qe_q(\bar{q}_jq_j)_{V+A},&
 O_8=\frac{3}{2}(\bar{s}_ib_j)_{V-A}\sum\limits_qe_q(\bar{q}_jq_i)_{V+A},
 \\
 O_9=\frac{3}{2}(\bar{s}_ib_i)_{V-A}\sum\limits_qe_q(\bar{q}_jq_j)_{V-A},&
 O_{10}=\frac{3}{2}(\bar{s}_ib_j)_{V-A}\sum\limits_qe_q(\bar{q}_jq_i)_{V-A}.
\end{array}
\end{equation}
where $q=s$. The first four operators are QCD penguin operators;
while the last four are electroweak penguin operators, which
should be suppressed by the coupling $\alpha/\alpha_s$.

 The decay width for this channel is:
\begin{eqnarray}
\Gamma=\frac{1}{2}\frac{G_F^2{\bf |P_c|}}{16\pi M_B^2}
|V_{tb}^*V_{td}|^2 \sum\limits_{\sigma =L,T}{\cal
M}^{\sigma\dag}{\cal M}^{\sigma} ,
\end{eqnarray}
where ${\bf P_c}$ is the 3-momentum of the final state meson, with
$|{\bf P_c}|=\frac{M_B}{2}(1-2r)$. Note that for our case an
additional factor $1/2$ should appear for the permutation symmetry
of the identical final state particles. The decay amplitude ${\cal
M}^{\sigma}$ which is decided by QCD dynamics will be calculated
later in PQCD approach. The subscript $\sigma$ denotes the
helicity states of the two vector mesons with L(T) standing for
the longitudinal (transverse) components. After analyzing the
Lorentz structure, the amplitude can be decomposed into \cite{1}:
\begin{eqnarray}
{\cal M}^{\sigma}=M_B^2{\cal M}_L+M_B^2{\cal
M}_N\epsilon^{\ast}_2(\sigma=T)\cdot \epsilon^{\ast}_3(\sigma=T)+
i{\cal
M}_T\epsilon_{\mu\nu\rho\sigma}\epsilon_2^{\mu\ast}
\epsilon_3^{\nu\ast}P_2^{\rho}P_3^{\sigma}.
\end{eqnarray}
We can define the longitudinal $H_0$, transverse $H_\pm$ helicity
amplitudes as
\begin{equation}
H_0=M^2_B{\cal M}_L,\,\,H_{\pm}=M^2_B{\cal M}_N\mp
M_{\phi}^2\sqrt{r^{\prime 2}-1}{\cal
M}_T,
\end{equation}
where $r^{\prime}=(P_2\cdot P_3)/ {M_{\phi}^2}$. After the
helicity summation, we can deduce that they satisfy the relation
\begin{equation}
\sum\limits_{\sigma =L,R}{\cal M}^{\sigma\dag}{\cal
M}^{\sigma}=|H_0|^2+|H_+|^2+|H_-|^2.
\end{equation}

There is  another equivalent set of definition of helicity
amplitudes
\begin{eqnarray}
\nonumber &&A_0=-\xi M^2_B{\cal M}_L,\\
\nonumber && A_{\|}=\xi \sqrt{2} M^2_B{\cal M}_N,\\
&& A_{\perp}=\xi M_{\phi}^2\sqrt{r^{\prime 2}-1}{\cal M}_T,
\end{eqnarray}
with $\xi$ the normalization factor to satisfy
\begin{eqnarray}
 |A_0|^2+|A_{\|}|^2+ |A_{\perp}|^2= 1,
\end{eqnarray}
where the notations $A_0$, $A_{\|}$, $A_{\perp}$ denote the
longitudinal, parallel, and perpendicular polarization amplitude.

What is followed is to calculate the matrix elements ${\cal M}_L$,
${\cal M}_N$ and ${\cal M}_T$ of various operators in the weak
Hamiltonian with PQCD approach. In PQCD approach, the decay
amplitude is factorized into the convolution of the mesons'
light-cone wave functions, the hard scattering kernel and the
Wilson coefficients, which stands for the soft, hard and harder
dynamics respectively. The transverse momentum was introduced so
that the endpoint singularity which will break the collinear
factorization is regulated and the large double logarithm term
appears after the integration on the transverse momentum, which is
then resummed into the Sudakov form factor. The formalism can be
written as:
\begin{eqnarray}
{\cal M}\sim\nonumber &&\int
dx_1dx_2dx_3b_1db_1b_2db_2b_3db_3Tr[C(t)\Phi_B(x_1,b_1)\Phi_{\phi}(x_2,b_2)
\Phi_{\phi}(x_3,b_3)\\ &&H(x_i,b_i,t)S_t(x_i)e^{-S(t)}],
\end{eqnarray}
where the $b_i$ is the conjugate space coordinate of the
transverse momentum, which represents the transverse interval of
the meson. $t$ is the largest energy scale in hard function $H$,
while the jet function $S_t(x_i)$ comes from the summation of the
double logarithms $\ln^2x_i$, called threshold resummation
\cite{18}, which becomes large near the endpoint.

The light cone wave functions of mesons are not calculable in
principal in PQCD, but they are universal for all the decay
channels. So that they can be constraint from the measured other
decay channels, like $B\to K\pi$ and $B\to \pi \pi$ decays etc.
\cite{10}. For the heavy $B$ meson, we have
\begin{eqnarray}
\frac{1}{\sqrt{2N_c}}(\not\!P_1+M_B)\gamma_5\phi_B(x,b).\label{13}
\end{eqnarray}
For the longitudinal polarized $\phi$ meson,
\begin{eqnarray}
\frac{1}{\sqrt{2N_c}}[M_{\phi}\not\!\epsilon_{2L}\phi_{\phi}(x)+\not\!\epsilon_{2L}
\not\!P_2\phi_{\phi}^t(x)+M_{\phi}I\phi^s_{\phi}(x)],
\end{eqnarray}
 and for transverse polarized $\phi$ meson,
\begin{eqnarray}
\frac{1}{\sqrt{2N_c}}[M_{\phi}\not\!\epsilon_{2T}\phi^v_{\phi}(x)+\not\!\epsilon_{2T}
\not\!P_2\phi_{\phi}^T(x)+\frac{M_{\phi}}{P_2\cdot
n_-}i\epsilon_{\mu\nu\rho\sigma}\gamma_5\gamma^{\mu}\epsilon_{2T}^{\nu}
P_2^{\rho}n_-^{\sigma}\phi^a_{\phi}(x)].
\end{eqnarray}
 In the following concepts, we omit the subscript of the $\phi$
meson for simplicity.
\begin{figure}[tbh]
\begin{center}
\epsfxsize=4.0in\leavevmode\epsfbox{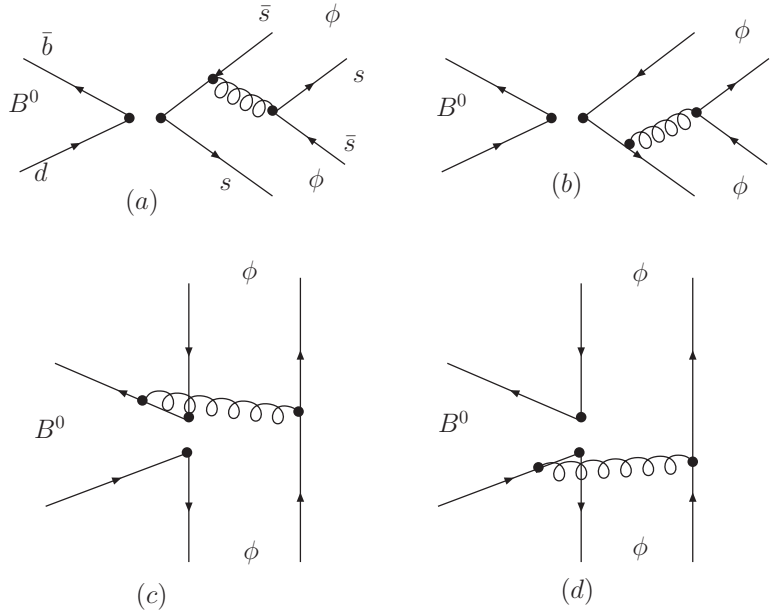}
\end{center}
\caption{{\protect\footnotesize Leading order Feynman diagrams for
$B^0\rightarrow \phi\phi$}}
\end{figure}

Now the only thing left is the hard part $H$. In PQCD approach, it
contains the corresponding four quark operator and the hard gluon
connecting the quark pair from sea. They altogether make an
effective six quark interaction. The hard part $H$ is channel
dependent, but it is perturbative calculable. When calculating the
hard parts (shown in the Figure 1), the factorizable diagrams (a)
and (b) have strong cancellation effects, which results in null
longitudinal polarization contribution and null parallel
polarization contribution. The perpendicular polarization survives
with a large factorizable contribution, which will be shown later
to make a large transverse polarization.  The detailed formulas
with polarization ${\cal M}_L$, ${\cal M}_N$, and ${\cal M}_T$ for
each diagram are given in the appendix. According to PQCD power
counting rules, the longitudinal nonfactorizable diagram should
give the leading contribution, and the contributions from the
other diagrams are suppressed by a factor $r$.

\section{Numerical results and summary}

\hspace*{\parindent} For the $B$ meson wave function distribution
amplitude in eq.(\ref{13}), we employ the model \cite{10}
\begin{eqnarray}
\phi_B(x)=N_Bx^2(1-x)^2\exp\left[-\frac{1}{2}\left(\frac{xM_B}{\omega_B}\right)^2
-\frac{\omega^2_Bb^2}{2}\right],
\end{eqnarray}
where the shape parameter $\omega_B=0.4$GeV has been constrained
in other decay modes. The normalization constant $N_B=91.784$GeV
is related to the B decay constant $f_B=0.19$GeV. It is one of the
two leading twist B meson wave functions; the other one is power
suppressed, so we omit its contribution in the leading power
analysis \cite{17}. The $\phi$ meson distribution amplitude up to
twist-3 are given by \cite{16}
\begin{eqnarray}
&&\phi_{\phi}(x)=\frac{3f_{\phi}}{\sqrt{2N_c}}x(1-x),\\
&&\phi^t_{\phi}(x)=\frac{f_{\phi}^T}{2\sqrt{2N_c}}\left\{3(1-2x)^2
+1.68C_4^{\frac{1}{2}}(1-2x)
+0.69\left[1+(1-2x)\ln{\frac{x}{1-x}}\right]\right\},\\
 &&\phi^s_{\phi}(x)=\frac{f_{\phi}^T}{4\sqrt{2N_c}}\left[3(1-2x)
 (4.5-11.2x+11.2x^2)+1.38\ln{\frac{x}{1-x}}\right],\\
&&\phi^T_{\phi}(x)=\frac{3f_{\phi}^T}{2\sqrt{2N_c}}x(1-x)
\left[1+0.2C_4^{\frac{3}{2}}(1-2x)
\right],\\
&&\phi^v_{\phi}(x)=\frac{f_{\phi}^T}{2\sqrt{2N_c}}\left\{\frac{3}{4}
[1+(1-2x)^2]+0.24[3(1-2x)^2-1]
+0.96C_4^{\frac{1}{2}}(1-2x)\right\},\\
&&\phi^a_{\phi}(x)=\frac{3f_{\phi}^T}{4\sqrt{2N_c}}(1-2x)[1+0.93(10x^2-10x+1)],
\end{eqnarray}
with the Gegenbauer polynomials,
\begin{eqnarray}
&&C_2^{\frac{1}{2}}(\xi)=\frac{1}{2}(3\xi^2-1),\\
&&C_4^{\frac{1}{2}}(\xi)=\frac{1}{8}(35\xi^4-30\xi^2+3),\\
&&C_2^{\frac{3}{2}}(\xi)=\frac{3}{2}(5\xi^2-1).
\end{eqnarray}

We employ the constants as follows \cite{pdg}: the Fermi coupling
constant $G_F=1.16639\times10^{-5}$GeV$^{-2}$, the CKM matrix
element $|V_{tb}^\ast V_{td}|=0.0084$, the meson masses
$M_B=5.28$GeV, $M_{\phi}=1.02$GeV, the decay constants
$f_{\phi}=0.237$GeV, $f_{\phi}^T=0.22$GeV and the $B$ meson
lifetime $\tau_{B^0}=1.55ps$. The results for the center value of
the branching ratio is then
\begin{eqnarray}
Br(B^0\rightarrow\phi\phi)=1.89\times 10^{-8},
\end{eqnarray}
and the helicity amplitudes are given by
\begin{eqnarray}
R_0=0.65,& R_{\|}=0.02, &R_{\perp}=0.33,
\end{eqnarray}
which shows that the transverse polarization contribution is
comparable to the longitudinal one.  The relative strong phases,
$\phi_{\|}=\arg{(A_{\|}/A_0)}$,
$\phi_{\perp}=\arg{(A_{\perp}/A_0})$ are given by
\begin{eqnarray}
\phi_{\|}=198.34^{\circ}, &\phi_{\perp}=195.48^{\circ}.
\end{eqnarray}

Now we consider the contribution from different operators. In the
factorizable diagrams, ${\cal M}_L={\cal M}_N =0$, because of the
cancellation between diagrams of Figure 1(a) and 1(b). For ${\cal
M}_T$, the QCD penguin operators $O_3$, $O_4$, $O_5$ and $O_6$,
contribute at the same level. In the nonfactorizable diagrams, the
operator $O_6$ give the most important contributions. If we omit
the contribution from the electroweak penguin operators, the
variation of the contribution from nonfactorizable diagrams
(Figure 1(c) and (d)) is small, while that of the factorizable
diagrams (Figure 1(a) and (b)) is large. The reason is that the
electroweak penguin operator $O_9$,  which has a large Wilson
coefficient, only presents in the factorizable diagrams. The
overall contribution of electroweak penguin at the branching ratio
level is less than $30\%$. We also test the contribution without
twist-3 wave functions. We find that if we keep only twist-2 wave
functions the total branching ratio doesn't change much, but the
contribution from the factorizable diagrams will vanish, and the
transverse polarization contribution then becomes very small. So
the twist-3 wave functions give very important corrections to the
polarization fractions.

There are many theoretical uncertainties in the calculation. The
next to leading order corrections to the hard part is a very
important kind of uncertainty for penguin dominant decays. To test
it, we consider the hard scale at a range
\begin{eqnarray}
\max(0.75M_BD_a,\frac{1}{b_2},\frac{1}{b_3})<t_a<\max(1.25M_BD_a,
\frac{1}{b_2},\frac{1}{b_3}),\\
 \max(0.75M_BD_b,\frac{1}{b_2},\frac{1}{b_3})<t_b<\max(1.25M_BD_b,
 \frac{1}{b_2},\frac{1}{b_3}),
\end{eqnarray}
\begin{eqnarray}
 \max(0.75M_BF,0.75M_BD_c, \frac{1}{b_1},\frac{1}{b_3})<t_c<\max(1.25M_BF,
 1.25M_BD_c, \frac{1}{b_1},\frac{1}{b_3}),\\
 \max(0.75M_BF, 0.75M_B|X|^{\frac{1}{2}},\frac{1}{b_1},\frac{1}{b_3})<
 t_d<\max(1.25M_BF, 1.25M_B|X|^{\frac{1}{2}},
\frac{1}{b_1},\frac{1}{b_3}),
\end{eqnarray}
and other parameters are fixed. Then we can obtain the value area
of the branching ratio as
\begin{eqnarray}
Br(B^0\rightarrow\phi\phi)=(1.89^{+0.61}_{-0.21})\times 10^{-8},
\end{eqnarray}
which is sensitive to the change of $t$, so the next to leading
order corrections will give important contribution. The ratios
$|A_0|^2$, $|R_{\|}|^2$ and $|R_{\perp}|^2$ are also very
sensitive to the variation of $t$, because that the nonfactorized
contributions decrease as the increasing of t, but the
factorizable diagrams, which gives the main contribution of the
transverse polarization, increase. The variety area of $|A_0|^2$
is about $0.41-0.81$ .

Another uncertainty is from the meson wave functions, which is
governed by other measured decays \cite{10}. The variation of the
parameters will also give the corrections, such as the parameter
$\omega_b$ in the $B$ wave function, if we assume its value area
is $0.32-0.48$, we will give the branching ratio
\begin{eqnarray}
Br(B^0\rightarrow\phi\phi)=(1.89^{+0.28}_{-0.26})\times 10^{-8}.
\end{eqnarray}
The ratios $R_0$, $R_{\|}$, $R_{\perp}$ is not very  sensitive to
the change of
 $\omega_b$, because it only gives an overall change of branching
 ratio, not to the individual polarization amplitudes.

In this paper, we calculate the rare decay channel
$B^0\rightarrow\phi\phi$ in PQCD approach and give its branching
ratio and polarization fractions in SM. This decay occur purely
via annihilation topology, and only penguin operators can
contribute. We predict that it has a very small branching ratio of
$10^{-8}$. This is so small that it will be sensitive to the new
physics, such as supersymmetry etc. \cite{7,yangyd}, which may
give a larger branching ratio. The current experiments only give
the upper limit: $Br(B^0\rightarrow \phi\phi)<1.5\times 10^{-6}$
\cite{19}, so the more accurate experimental results are needed to
test the theory.

\section*{Acknowledgments}

This work is partly supported by the National Science Foundation
of China under Grant No.90103013, 10475085  and 10135060, Y-L Shen
and J. Zhu thank Y. Li, and X-Q Yu for the help on the program,
Y-L Shen also thanks J-F Cheng and M-Z Yang for helpful
discussions.

\begin{appendix}

\section{factorization formulas}

In the factorizable diagrams, due to the identical particles at
the final states cancellation occurs between the two diagrams
figure (a) and (b). Only the perpendicular polarization part
survives,
\begin{eqnarray}
 {\cal M}_{T}^a \nonumber &=&-16\pi C_F f_BM_B^2 \int^1_0 dx_2
dx_3\int ^{\infty}_0b_2db_2\int
^{\infty}_0b_3db_3r\alpha_s(t)\left[\phi^v(x_2)\phi^v(x_3)(x_3-1)\right.\\
\nonumber &&+ \left.\phi^v(x_2)\phi^a(x_3)(1+x_3)
+\phi^a(x_2)\phi^v(x_3)(1+x_3)
+\phi^a(x_2)\phi^a(x_3)(x_3-1)\right]S_{\phi}(t)^2\\
&&\left[C_3+\frac{C_4}{3}-C_5-\frac{C_6}{3}+\frac{1}{2}(C_7+\frac{C_8}{3})
-\frac{1}{2}(C_9+\frac{C_{10}}{3})\right](t) h(x_2,x_3,b_2,b_3),
\\
 {\cal M}_{T}^b \nonumber &=&16\pi
C_F f_BM_B^2 \int^1_0 dx_2 dx_3\int ^{\infty}_0b_2db_2\int
^{\infty}_0b_3db_3r\alpha_s(t)[\phi^v(x_2)\phi^v(x_3)(x_2)\\
\nonumber &&+ \phi^v(x_2)\phi^a(x_3)(2-x_2)
+\phi^a(x_2)\phi^v(x_3)(2-x_2)
+\phi^a(x_2)\phi^a(x_3)(x_2)]S_{\phi}(t)^2\\
&&\left[C_3+\frac{C_4}{3}-C_5-\frac{C_6}{3}+\frac{1}{2}(C_7+\frac{C_8}{3})
-\frac{1}{2}(C_9+\frac{C_{10}}{3})\right](t)
h^{\prime}(x_2x_3,b_2,b_3),
\end{eqnarray}
where the $h$ functions come from the integral on the transverse
momentum, their manifest forms is
\begin{eqnarray}
\nonumber h(x_2,x_3,b_2,b_3)&=&(\frac{i\pi}{2})^2H_0^1(M_BF
b_2)S_t(x_3)[\theta(b_3-b_2) J_0(b_2M_BD_a)H_0^1(b_3M_BD_a)
\\
&&+\theta(b_2-b_3)J_0(b_3M_BD_a) H_0^1(b_2M_BD_a)],\\ \nonumber
h^{\prime}(x_2,x_3,b_2,b_3)&=&(\frac{i\pi}{2})^2H_0^1(M_BF
b_3)S_t(x_2)[\theta(b_3-b_2) J_0(b_2M_BD_b)H_0^1(b_3M_BD_b)
\\
&&+\theta(b_2-b_3)J_0(b_3M_BD_b)H_0^1(b_2M_BD_b)],
\end{eqnarray}
with the notation F and D stand for:
\begin{eqnarray}
\nonumber&&F=\sqrt{[(1-x_2)(1-r)+x_3r][x_3(1-r)+(1-x_2r)]}\\
\nonumber&&D_a=\sqrt{[x_3+r(1-x_3)][1-r(1-x_3)]}\\
&&D_b=\sqrt{[1-x_2+rx_2](1-rx_2)}.
\end{eqnarray}
$t$ is the hard scale, which is chosen as
\begin{eqnarray}
t_a=\max(M_BD_a, 1/b_2, 1/b_3),
 &&t_b=\max(M_BD_b, 1/b_2, 1/b_3).
\end{eqnarray}
The Sudakov form factor is written as
\begin{eqnarray}
S_{\phi}(t)=\exp{\left[-s(x_2P_2^+,b_2)-s((1-x_2)P_2^+,b_2)-2\int^t_{1/{b_2}}
\frac{d\bar{\mu}}{\bar{\mu}}\gamma(\alpha_s(\bar{\mu}^2))\right]},
\end{eqnarray}
with the quark anomalous dimension $\gamma=-{\alpha_s}/{\pi}$ and
the $s(Q,b)$, the so-called Sudakov factor, which comes from the
resummation of the double logarithms, is given as
\begin{eqnarray}
  s(Q,b) &=& \int_{1/b}^Q \!\! \frac{d\mu'}{\mu'} \left[
 \left\{ \frac{2}{3}(2 \gamma_E - 1 - \log 2) + C_F \log \frac{Q}{\mu'}
 \right\} \frac{\alpha_s(\mu')}{\pi} \right. \nonumber \\
& &  \left.+ \left\{ \frac{67}{9} - \frac{\pi^2}{3} -
\frac{10}{27} n_f
 + \frac{2}{3} \beta_0 \log \frac{\gamma_E}{2} \right\}
 \left( \frac{\alpha_s(\mu')}{\pi} \right)^2 \log \frac{Q}{\mu'}
 \right].
 \label{eq:SudakovExpress}
\end{eqnarray}

The nonfactorizable amplitudes for diagrams (c) and (d) are
written as
\begin{eqnarray}
M_{L}^c &=&\nonumber -\frac{32\pi C_FM_B^2 }{\sqrt{6}}
\int[dx]\int^{\infty}_0b_1db_1b_3db_3\phi_B(x_1)\{[-1+x_2+r(2-4x_2)]\\
\nonumber
&&\phi(x_2)\phi(x_3)+r(1+x_2-x_3)\phi^t(x_2)\phi^t(x_3)+r(-1+x_2+x_3)
\phi^t(x_2)\phi^s(x_3)\\ \nonumber &&+r(1-x_2-x_3)
\phi^s(x_2)\phi^t(x_3)+r(3-x_2+x_3) \phi^s(x_2)\phi^s(x_3)\}
\alpha_s(t)\\
&&[C_4+C_6- C_8/2-  C_{10}/2]
h_n(x_1,x_2,x_3,b_1,b_3) S(t),
\\
M_{L}^d&=&\nonumber -\frac{32\pi
C_FM_B^2}{\sqrt{6}}\int^1_0[dx]\int^{\infty}_0b_1db_1b_3db_3
\phi_B(x_1)\{[x_3-4x_3r]\phi(x_2)\phi(x_3)\\
\nonumber &&+r(1-x_2+x_3)\phi^t(x_2)\phi^t(x_3)
-r(1-x_2-x_3)\phi^t(x_2)\phi^s(x_3)\\ \nonumber
&&+r(1-x_2-x_3)
\phi^s(x_2)\phi^t(x_3)+r(-1+x_2-x_3)\phi^s(x_2)\phi^s(x_3)\}
\alpha_s(t)\\&&[C_4+C_6- C_8/2- C_{10}/2](t)
h_n^{\prime}(x_1,x_2,x_3,b_1,b_3)S(t),
\end{eqnarray}
\begin{eqnarray}
 {\cal M}_{N}^c \nonumber &=&-\frac{32\pi
C_FM_B^2 }{\sqrt{6}}
\int[dx]\int^{\infty}_0b_1db_1b_3db_3\phi_B(x_1)r[-2\phi^v(x_2)\phi^v(x_3)\\
\nonumber &&+ \phi^T(x_2)\phi^T(x_3)(1+x_2-x_3)
-2\phi^a(x_2)\phi^a(x_3)]\alpha_s(t)\\
&&\left[C_4+C_6-
C_8/2-C_{10}/2\right](t)h_n(x_1,x_2,x_3,b_1,b_3)S(t),
\\
 {\cal M}_{N}^d \nonumber &=&-\frac{32\pi
C_FM_B^2 }{\sqrt{6}}
\int[dx]\int^{\infty}_0b_1db_1b_3db_3\phi_B(x_1)r
\phi^T(x_2)\phi^T(x_3)(1-x_2\\
&&+x_3) \alpha_s(t)[C_4+C_6-
C_8/2-C_{10}/2](t)h_n^{\prime}(x_1,x_2,x_3,b_1,b_3)S(t),
\\
 {\cal M}_{T}^c \nonumber &=&\frac{64\pi
C_FM_B^2 }{\sqrt{6}}
\int[dx]\int^{\infty}_0b_1db_1b_3db_3\phi_B(x_1)r[2\phi^v(x_2)\phi^a(x_3)\\
\nonumber && +\phi^T(x_2)\phi^T(x_3)(1-x_2-x_3)
+2\phi^a(x_2)\phi^v(x_3)]\alpha_s(t)\\
&& [C_4-C_6+C_8/2- C_{10}/2](t)h_n(x_1,x_2,x_3,b_1,b_3)S(t),
\\
 {\cal M}_{T}^d \nonumber &=&-\frac{64\pi
C_FM_B^2 }{\sqrt{6}}
\int[dx]\int^{\infty}_0b_1db_1b_3db_3\phi_B(x_1)r
\phi^T(x_2)\phi^T(x_3)(1-x_2\\
&&-x_3) \alpha_s(t)[C_4-C_6+C_8/2-
C_{10}/2](t)h_n^{\prime}(x_1,x_2,x_3,b_1,b_3)S(t).
\end{eqnarray}
The $h$ functions are defined as
\begin{eqnarray}
h_n(x_1,x_2,x_3,b_1,b_3)
 \nonumber &=&\frac{i\pi}{2}\left[\theta(b_3-b_1)J_0(b_1M_BF) H_0^1(b_3M_BF)
 +\theta(b_1-b_3)\right.\\ && \left.J_0(b_3M_BF)
H_0^1(b_1M_BF)\right]K_0(M_BD_c b_1),\\
h_n^{\prime}(x_1,x_2,x_3,b_1,b_3)\nonumber &
=&\frac{i\pi}{2}\left[\theta(b_3-b_1)J_0(b_1M_BF) H_0^1(b_3M_BF)
 +\theta(b_1-b_3) \right.\\ &&\left.J_0(b_3M_BF)
 H_0^1(b_1M_BF)\right]
\times \left\{\begin{array}{cc} \frac{i\pi}{2}H_0^{(1)}(\sqrt{-X}
b_1), \hspace{0.3cm}&X<0,
\\ K_0(\sqrt{X}b_1),\hspace{0.3cm}&X>0,
\end{array}\right.
\end{eqnarray}
with the notations
\begin{eqnarray}
&&\int[dx]=\int_0^1dx_1\int_0^1dx_2\int_0^1dx_3\\ &&D_c
=\sqrt{1-[x_2-x_1+r(1-x_2-x_3)][1-x_3-r(1-x_2-x_3)]}\\
&&
 X=[x_2+x_1-1+r(1-x_2-x_3)][x_3+r(1-x_2-x_3)].
\end{eqnarray}
And the hard scale $t$ is
\begin{eqnarray}
 &&t_c=\max(M_BF,M_BD_c, 1/b_1, 1/b_3),\\&& t_d=\max(M_BF, M_B\sqrt{|X|},
1/b_1, 1/b_3).
\end{eqnarray}
The Sudakov form factor is $S(t)=S_B(t)S^2_{\phi}(t)$, with
\begin{eqnarray}
S_B(t)=s(x_1P_1^+,b_1)+2\int^t_{1/{b_1}}
\frac{d\bar{\mu}}{\bar{\mu}}\gamma(\alpha_s(\bar{\mu}^2)).
\end{eqnarray}
\end{appendix}

\end{document}